\documentclass[12pt]{iopart}

\usepackage{times}

\newcommand{\be}{\begin{equation}}
\newcommand{\ee}{\end{equation}}
\newcommand{\ba}{\begin{eqnarray}}
\newcommand{\ea}{\end{eqnarray}}

\newcommand{\nn}{\nonumber\\}

\begin{document}

\title{ Electromagnetic response and effective
gauge theory of double-layer quantum Hall systems}
 \author{K. Shizuya}
  \address{Yukawa Institute for Theoretical Physics\\
 Kyoto University,~Kyoto 606-8502,~Japan }

\begin{abstract} 
We report on an effective gauge theory of double-layer quantum Hall
systems, that is constructed via bosonization from the response
of incompressible states without referring to composite bosons and
fermions. It is pointed out that dipole-active excitations, both
elementary and collective, govern the long-wavelength features of quantum
Hall systems, and the single-mode approximation is used to study them.
The effective theory consists of three vector
fields representing  one interlayer collective mode and two cyclotron
modes, and properly incorporates the spectrum of collective excitations
on the right scale of the Coulomb interaction. 
Special emphasis is placed on exploring the advantage of looking into
quantum Hall systems through their response; in particular, subtleties
inherent to the standard Chern-Simons theories are critically examined.
\end{abstract}



\section{Introduction}

The Chern-Simons (CS) theories~\cite{ZHK,LF} realize the
composite-boson and composite-fermion pictures~\cite{GM,J} of the
fractional  quantum Hall effect (FQHE) and have been successful in
describing various features of the fractional quantum Hall states.  
They, however, have some subtle limitations as well~\cite{MZ}.  
In particular, when adapted to double-layer systems~\cite{BIH,CP,YMG},
they differ significantly in collective-excitation spectrum from the 
magneto-roton theory of Girvin, MacDonald and Platzman~\cite{GMP}, based
on the single-mode approximation (SMA).

In this paper we wish to report on a new approach~\cite{KS} to effective
theories of the FQHE, that relies on the incompressible nature of the
quantum Hall states without referring to the composite-boson or
composite-fermion picture.  The basic tools used are
projection to the Landau levels and functional bosonization~\cite{FS}, which
are combined to construct, via the electromagnetic response of the
incompressible states, a long-wavelength effective theory of the FQHE.  
For single-layer systems this effective theory
properly reproduces the results consistent with the CS theories.
Our approach thus has logic independent of but complementary to the
standard CS theories. 

The real merit of our approach becomes prominent for double-layer
systems. We shall examine the electromagnetic
characteristics of double-layer systems within the SMA theory and derive
an effective theory that  properly incorporates the SMA spectrum of
collective excitations. We shall thereby clarify the relation between the
SMA theory and the CS theories. The effects of interlayer
coherence and tunneling are also discussed.

\section{Double-layer systems}

Consider first a double-layer system in the absence of interlayer tunneling,
with  average electron densities 
$\rho_{\rm av}^{(\alpha)}=(\rho_{\rm av}^{(1)},\rho_{\rm av}^{(2)})$ in the
upper $(\alpha =1)$ and lower $(\alpha =2)$ layers.  
The electrons in the two layers are coupled through the
Coulomb interaction $V^{\alpha\beta}_{\bf p}$, where 
$V^{11}_{\bf p}=V^{22}_{\bf p}$ and 
$V^{12}_{\bf p} = V^{21}_{\bf p}$ denote the intralayer and interlayer
potentials, respectively.

The system is placed in a common strong perpendicular magnetic field
$B_{z}=B>0$. We suppose that the electron fields 
$\psi^{(\alpha)}$ in each
layer are fully spin polarized and  
assemble them into a pseudospin~\cite{Moon} doublet spinor 
$\Psi=(\psi^{(1)},\psi^{(2)})^{\rm tr}$. 
Our task is to study how the system responds to weak electromagnetic potentials
$A^{(\alpha)}_{\mu}(x)$ $(\mu=0,1,2)$ in three space, and we thus write
the one-body Hamiltonian in the form
\begin{eqnarray}
H_{1} =  \int d^{2}{\bf x}\,  \Psi^{\dag}\, \left[
 {1\over{2M}}({\bf p} \!+ \! {\bf A}^{\!B}\!+ \! {\bf A}^{\!+}\!
+ {\bf A}^{\!-}\sigma_{3})^{2}
 +A_{0}^{\!+} +A_{0}^{\!-} \sigma_{3} \right] \Psi,
\label{Hone}
\end{eqnarray}
where
$A_{\mu}^{\pm}(x)= {1\over{2}}\,\{A^{(1)}_{\mu}(x) \pm A^{(2)}_{\mu}(x)\}$
in terms of the potentials acting on each layer; 
${\bf A}^{\!B}=eB\,(-y,0)$ supplies a uniform magnetic field $B$; 
the electric charge $e>0$ has been suppressed by rescaling $eA_{\mu}\rightarrow
A_{\mu}$.  The $A_{0}^{-}$ coupled to $\rho^{(1)} - \rho^{(2)}$ probes 
out-of-phase density fluctuations of the two layers while $A_{0}^{+}$
probes in-phase density fluctuations.

Let us now project our system into Landau levels.   
Let $|N\rangle =|n,y_{0}\rangle$ denote  the Landau levels of a freely
orbiting electron of energy $\omega_{c} (n+{\scriptstyle {1\over2}})$
with $n = 0,1,2,\cdots$, and $y_{0}=\ell^{2}\,p_{x}$, where
$\omega_{c} \equiv eB/M$ and ${\ell}\equiv 1/\sqrt{eB}$; 
we set ${\ell} \rightarrow 1$ below. 
We first pass into $N=(n,y_{0})$ space by a unitary transformation
$\Psi ({\bf x},t)=\sum_{N} \langle {\bf x}|N\rangle\, \Phi_{n}(y_{0},t)$ and
then, by a subsequent unitary transformation 
$\Phi_{n}(y_{0},t) \rightarrow \Psi_{m}(y_{0},t)$, make the one-body Hamiltonian
diagonal in level indices. The resulting projected Hamiltonian is an operator in 
${\bf r} \equiv (r_{1},r_{2}) = (i\ell^{2} \partial/\partial y_{0}, y_{0})$ with
uncertainty $[r_{1}, r_{2}]=i\ell^{2}$.  

Let us focus on the lowest Landau level $n=0$ in a strong magnetic field.
The projected one-body Hamiltonian to $O(A^{2})$ reads 
$\bar{H}^{\rm cyc}+ \bar{H}^{\rm em}$ with~\cite{KS}
\begin{eqnarray}
\bar{H}^{\rm cyc} &=& \sum_{\alpha=1}^{2}
\sum_{\bf p}\Big\{{\omega_{c}\over{2}}\, \delta_{\bf p,0}
+{\cal U}^{(\alpha)}_{\bf p}\Big\} \,
\bar{\rho}_{\bf -p}^{(\alpha)} ,\ \ 
\bar{H}^{\rm em} = \sum_{\bf p}\Big\{ \chi^{+}_{\bf p} \,
\bar{\rho}_{\bf -p} +2 \chi^{-}_{\bf p}\, \bar{S}^{3}_{\bf -p}\Big\} ,
\end{eqnarray}
where $\chi_{\bf p}^{\pm} =(A_{0}^{\pm})_{\bf p} +
(1/2M)({A}_{12}^{\pm})_{\bf p}$; $(A_{\mu}^{\pm})_{\bf p}$ stands for the Fourier
transform of $A_{\mu}^{\pm}(x)$, $A_{12}^{\pm} =
\partial_{1}A_{2}^{\pm} - \partial_{2}A_{1}^{\pm}$.
Here the projected charges $\bar{\rho}_{\bf p}=
\bar{\rho}^{(1)}_{\bf p} +
\bar{\rho}^{(2)}_{\bf p}$, $\bar{S}^{3}_{\bf p}=
{1\over{2}}\,(\bar{\rho}^{(1)}_{\bf p} -\bar{\rho}^{(2)}_{\bf p})$, etc., are
defined by 
\ba
 \left(\bar{\rho}_{\bf p}, \bar{S}^{a}_{\bf p}\right) &\equiv&
\int dy_{0}\ \Psi_{0}^{\dag}(y_{0},t)\,
e^{-{1\over{4}}\, {\bf p}^{2}}\, e^{-i{\bf p\cdot r}}\,
 \left(1, {1\over{2}} \sigma_{a}\right)\Psi_{0}(y_{0},t), 
\end{eqnarray}
where the two-spinor $\Psi_{0}$ defines the true lowest
Landau level.
The ${\cal U}^{(\alpha)}_{\bf p}$ denote the
contributions quadratic in $A_{\mu}^{(\alpha)}$, and are given (for ${\bf p}=0$)  by
${\cal U}_{\bf p=0}= \int d^{2}{\bf x}\, {\cal U}(x)$  with
\begin{eqnarray}
{\cal U}^{(\alpha)}(x) 
&=& {1\over{2}}\, A^{(\alpha)}_{\mu} D \epsilon^{\mu \nu \rho}\partial_{\nu}A^{(\alpha)}_{\rho}
-{1\over{2\omega_{c}}} A^{(\alpha)}_{k0} D A^{(\alpha)}_{k0}  +\cdots, \label{Utwo}
\end{eqnarray}
where $D=\omega_{c}^{2}/(\omega_{c}^{2} + \partial_{t}^{2} )$ and 
$A_{\mu \nu}= \partial_{\mu}A_{\nu}- \partial_{\nu}A_{\mu}$; 
$\epsilon^{\mu\nu\rho}$ is a totally-antisymmetric tensor with 
$\epsilon^{012} =1$. 

The charges $(\bar{\rho}_{\bf p},\bar{S}^{a}_{\bf p})$ obey 
a simple SU(2)$\times W_{\infty}$ algebra~\cite{Moon} 
\begin{eqnarray}
&&[\bar{\rho}_{\bf p}, \bar{\rho}_{\bf k}]
= c(p,k) \, \bar{\rho}_{\bf p+k} , \ \ \
[\bar{\rho}_{\bf p}, \bar{S}^{a}_{\bf k}]
= c(p,k) \, \bar{S}^{a}_{\bf p+k}, \nonumber\\
&&[\bar{S}^{a}_{\bf p},\bar{S}^{b}_{\bf k}]= 
h(p,k)\,i\epsilon^{abc}\, \bar{S}^{c}_{\bf p+k}
+\delta^{ab}\, {1\over{4}}\,  c(p,k)\,\bar{\rho}_{\bf p+k},
\label{chargealgebra}
\end{eqnarray}
with some given functions $c(p,k)$ and $h(p,k)$. 
They, however, differ slightly~\cite{KS} from the projected charge
densities 
$(\rho_{00}^{(\alpha)})_{\bf p}= \bar{\rho}^{(\alpha)}_{\bf p} +
\triangle \bar{\rho}^{(\alpha)}_{\bf p}$.
The corrections $\triangle \bar{\rho}^{(\alpha)}_{\bf p}$ depend on the
potentials $A^{(\alpha)}_{\mu}$ and, as a result, the projected Coulomb
interaction
\begin{eqnarray}
\bar{H}^{\rm Coul}
=&&{1\over{2}}\, \sum_{\bf p} \Big( V^{+}_{\bf p}\,
\bar{\rho}_{\bf- p}\,\bar{\rho}_{\bf p} +
4V^{-}_{\bf p}\, \bar{S}^{3}_{\bf- p}\, \bar{S}^{3}_{\bf p} \Big)
+ \triangle H^{\rm Coul}
\label{projCoul}
\end{eqnarray}
acquires a field-dependent piece $\triangle H^{\rm Coul}$, which plays an
important role, as we shall see.  
Here we have set $V^{\pm}_{\bf p}\equiv {\scriptstyle {1\over{2}}}\, (V^{11}_{\bf p} \pm 
V^{12}_{\bf p})$.

The dynamics within the lowest Landau level is now governed by the
Hamiltonian $\bar{H}= \bar{H}^{\rm Coul} +\bar{H}^{\rm cyc} +\bar{H}^{\rm em}$.
Suppose now that an incompressible many-body state $|G\rangle$ of
uniform density $(\rho_{\rm av}^{(1)},\rho_{\rm
av}^{(2)})$ is formed within the lowest Landau level. Then, setting 
$\langle G| \bar{\rho}^{(\alpha)}_{\bf -p} |G \rangle 
=\rho^{(\alpha)}_{\rm av}\, (2\pi)^{2}\delta^{2} ({\bf p})$ 
in $\bar{H}^{\rm em}$ one obtains the effective action to $O(A^{2})$:
\begin{eqnarray}
S^{\rm cycl}= -\int dtd^{2}{\bf x} \sum_{\alpha}\rho_{\rm
av}^{(\alpha)}\,  {\cal U}_{2}^{(\alpha)}(x),
\label{Sem}
\end{eqnarray}
which summarizes the response due to 
electromagnetic inter-Landau-level mixing, i.e., due to the cyclotron
modes (one for each layer).

The electromagnetic interaction in $\bar{H}$ also gives rise to
intra-Landau-level transitions.
For single-layer systems the intra-Landau-level excitations  are only
dipole-inactive~\cite{GMP}  (i.e., the response vanishes faster than
${\bf k}^{2}$ for ${\bf k} \rightarrow 0$) as a result of Kohn's theorem,
and the incompressible quantum Hall states show universal $O({\bf k})$
and $O({\bf k}^{2})$ long-wavelength electromagnetic characteristics
determined by the cyclotron mode alone. The situation changes drastically for
double-layer systems, which we discuss in the next section.

\section{Electromagnetic response}

In this section we examine the electromagnetic response of
double-layer  systems due to intra-Landau-level processes by
use of the projected single-mode approximation (SMA)~\cite{GMP}.
Let $|G\rangle$ denote the exact ground state of the present double-layer
system.
The SMA supposes that the density fluctuations over $|G\rangle$ have
predominant overlap with $|G\rangle$ through the associated density
operators $\rho^{(\alpha)}$.  We here consider two modes, a phonon-roton
mode $|\phi^{-}_{\bf k}\rangle \sim \bar{S}^{3}_{\bf k}|G\rangle$
representing the out-of-phase density fluctuations  of the two layers and
a charge mode $|\phi^{+}_{\bf k}\rangle \sim \bar{\rho}_{\bf k}|G\rangle$ 
representing the in-phase density fluctuations.

The normalization of the $|\phi^{-}_{\bf k}\rangle$ mode
\begin{eqnarray}
\bar{s}^{-}({\bf k}) &\equiv& {2\over{N_{e}}}\,
\langle G|\bar{S}^{3}_{\bf -k}\,\bar{S}^{3}_{\bf k}\, |G\rangle 
=\bar{s}^{-}({\bf -k})
\end{eqnarray}
is called the (projected) static structure factor of the
ground state $|G\rangle$; $N_{e}=N_{e}^{(1)}+N_{e}^{(2)}$.
Similarly, we normalize $|\phi^{+}_{\bf k}\rangle$
with $\bar{s}^{+}({\bf k}) =\langle G|
\bar{\rho}_{\bf -k}\,\bar{\rho}_{\bf k}\, |G\rangle /(2N_{e})$.
In the SMA the static structure factors $\bar{s}^{\pm}({\bf k})$
are the basic quantities, through which the effect of nontrivial
correlations pertinent to the ground state is reflected in
the dynamics. 
Invariance under translations of both layers implies an analog of
Kohn's theorem for the interlayer in-phase collective excitations so
that $\bar{s}^{+}({\bf k}) \sim |{\bf k}|^{4}$ for small 
${\bf k}$~\cite{MZ,RR}. As a result, the $O({\bf k})$ and $O({\bf k}^{2})$
in-phase response of the double-layer system is
governed by the cyclotron modes.

On the other hand, the presence of interlayer interactions 
$V^{12}_{\bf p}$ spoils invariance under relative translations of the
two layers and, unless spontaneous interlayer coherence is realized,
the out-of-phase collective excitations become
dipole-active~\cite{MZ,RR},
\begin{equation}
\bar{s}^{-}({\bf k}) =
c^{-}\,{1\over{2}}\,{\bf k}^{2} +  O(|{\bf k}|^{4}).
\label{sminus}
\end{equation}
With this fact in mind we shall henceforth concentrate on the
dipole-active out-of-phase response. 
Actually,  particular sets of double-layer QH states of our concern  are
the Halperin  $(m_{1},m_{2},n)$ or $(m,m,n)$  states~\cite{BIH}.
For the $(m,m,n)$ states the coefficient $c^{-}$ is 
known~\cite{RR}:
\begin{eqnarray}
c^{-} = {n/(m-n)} .
\label{cminus}
\end{eqnarray}

To determine the excitation energy $\epsilon^{-}_{\bf k}$ of
the collective mode in the SMA one considers the
(projected) oscillator strength
$\bar{f}^{-}({\bf k}) = (2/N_{e})\, \langle G|\,
\bar{S}^{3}_{\bf -k}\, [\bar{H} , \bar{S}^{3}_{\bf k}]\, |G\rangle$, 
which is calculable~\cite{MZ,RR} by use of the
algebra~(\ref{chargealgebra}).
Saturating it with the single mode $|\phi^{-}_{\bf k}\rangle$ then 
yields the SMA excitation gap $\epsilon^{-}_{\bf k} = \bar{f}^{-}({\bf
k})/\bar{s}^{-}({\bf k})$ at  ${\bf k}\rightarrow 0$:
\begin{equation}
\epsilon^{-}_{0}\equiv \epsilon^{-}_{\bf k=0}
={1\over{c^{-}}}\, \sum_{\bf p}V^{12}_{\bf p}\,{\bf p}^{2}
\{-\bar{s}^{12}({\bf p})\},
\label{eminuszero}
\end{equation}
where $ \bar{s}^{12} ({\bf p}) 
={1\over{2}}\, \{ \bar{s}^{+} ({\bf p}) -\bar{s}^{-} ({\bf p}) \}$.

Let us now study the response of the state $|G\rangle$ through
the collective excitations. Consider first the $A^{-}_{0}$ to $A^{-}_{0}$
response, and write the associated density-density response function 
$(-4i)\langle G|T\bar{S}^{3}(x)\bar{S}^{3}(x') |G\rangle$ in spectral form
in Fourier space
\begin{eqnarray}
F[\omega,{\bf k}]
&=&\sum_{n}\Big\{{1\over{\omega  -\epsilon_{n}}}\, \sigma^{n}({\bf k})
-{1\over{\omega + \epsilon_{n}}}\, \sigma^{n}({\bf -k}) \Big\}.
\label{spectralform}
\end{eqnarray}
In the SMA we saturate the sum over $|n\rangle$ by a single collective
mode $|\phi^{-}_{\bf k}\rangle$
so that the spectral weight 
$\sigma^{n}({\bf k})= (4/\Omega)\,
\langle G|\, \bar{S}^{3}_{\bf k}\,  |n\rangle
\langle n|\,\bar{S}^{3}_{\bf -k} |G\rangle  \rightarrow
(2N_{e}/\Omega)\,\bar{s}^{-}({\bf k})$, where $\Omega$ denotes the spatial
surface area. With $\bar{s}^{-}({\bf k})\approx {1\over{2}}c^{-}{\bf
k}^{2}$, this leads to a dipole response of the form
\begin{equation}
S^{\rm col}_{A_{0}} =  {1\over{2}}\, \rho_{\rm av}
\int d^{3}x\,  \partial_{j}A^{-}_{0}(x) \,
{2\, c^{-}\,\epsilon^{-}_{0}\over{ 
(\epsilon^{-}_{0})^{2}} \!-\!\omega^{2}}\,
\partial_{j}A^{-}_{0}(x) ,
\label{AzeroAzero}
\end{equation}
where $\omega$ stands for $i\partial_{t}$ and  $\rho_{\rm av} = \rho^{(1)}_{\rm
av} + \rho^{(2)}_{\rm av}$.
This response is not gauge invariant by itself. Fortunately, the
field-dependent Coulomb interaction $\triangle H^{\rm Coul}$ serves to
promote it into a gauge-invariant form and at the same time yields
another important response, the Chern-Simons term.
The actual calculation is quite nontrivial since it requires evaluation of
products of three charge operators such as $
\langle G|\, \bar{S}^{3}_{\bf k}\, \{\bar{\rho}_{\bf -p},\bar{S}^{3}_{\bf
p-k}\}|G\rangle $. Nevertheless, it turns out that the leading
long-wavelength response is determined from the portion that derives from
the noncommutative nature $([r_{1},r_{2}]\not=0)$ of the projected
charges.  Eventually one is led to the dipole and related  response of the
form~\cite{KS}
\begin{eqnarray}
S_{\rm eff}^{-}
&=& {\rho_{\rm av}\over{2}}\int\! d^{3}x\, A^{-}_{j0}\, \Big[
{ 2\, c^{-}\,\epsilon^{-}_{0}\over{ (\epsilon^{-}_{0})^{2}}
\!-\! \omega^{2}}\, +
{\omega_{c}\over{ \omega _{c}^{2}} \!-\! \omega^{2}}\,\Big]
A^{-}_{j0} 
\nonumber\\  && 
- {\rho_{\rm av}\over{2}} \int\! d^{3}x\, A^{-}_{\mu} \,
\Big[{ 2 c^{-}\,(\epsilon_{0}^{-})^{2}
\over{(\epsilon^{-}_{0})^{2}-\omega^{2}}}
+{\omega_{c}^{2}\over{ \omega _{c}^{2}} \!-\!
\omega^{2}}\,\Big]\,\epsilon^{\mu \nu\rho}
\partial_{\nu}A^{-}_{\rho} 
\nonumber\\ && 
-{\rho_{\rm av}\over{2M}} \int\! d^{3}x\, A_{12}^{-}\,
\Big[\,
{ 2\, c^{-}\,(\epsilon^{-}_{0})^{2}
\over{(\epsilon^{-}_{0})^{2}} \!-\! \omega^{2}}\, +
{\omega_{c}^{2}\over{ \omega _{c}^{2}} \!-\! \omega^{2}}\,\Big]
 A_{12}^{-},
\label{fullresponse}
\end{eqnarray}
where  the cyclotron-mode contribution in Eq.~(\ref{Sem}) has also been
included.

The effect of the collective mode is significant in the low-energy
$(\omega \ll \epsilon^{-}_{0})$ regime. 
The dipole response acquires the scale $\epsilon^{-}_{0}$ rather
than $\omega_{c}$.  The Hall conductance and magnetic susceptibility are
enhanced by a common factor $(2 c^{-}+1)$, which equals $(m+n)/(m-n)$ for
the $(m,m,n)$ states. 
Actually, the Hall conductance $\sigma^{-}_{xy}=-e^{2}\rho_{\rm
av}\,(m+n)/(m-n)$ and the static density correlation function
$\rho _{\rm av} (2c^{-}+1)\, {\bf p}^{2}$  agree exactly with 
what one gets in the Chern-Simons theories~\cite{WZdlayer,EI,LFdl}.

\section{Effective gauge theory}
 
In the previous section we have calculated the response of a
double-layer electron system. Once such a response is known, it is
possible to reconstruct it through the quantum fluctuations of a boson
field, a procedure known as functional bosonization~\cite{FS}. 
Indeed, it is a simple exercise of functional integration to
verify that, of the out-of-phase response $S_{\rm eff}^{-}$ in
Eq.~(\ref{fullresponse}),  the collective-mode contribution is
reconstructed from the theory of a three-vector field
$\xi_{\mu}=(\xi_{0},\xi_{1},\xi_{2})$, with the Lagrangian~\cite{KS}
\begin{eqnarray}
L^{\rm col}_{\xi}&=& -A_{\mu}^{-} \epsilon^{\mu \nu\rho}
\partial_{\nu}\xi_{\rho}
+ {1\over{4 c^{-}\rho_{\rm av}}}
\Big[ \xi_{\mu} \epsilon^{\mu\nu\rho}\partial_{\nu} \xi_{\rho}
+ {1\over{\epsilon_{0}^{-}}}\,(\xi_{k0})^{2}
- {1\over{M}}\, (\xi_{12})^{2}\Big].
\label{Lxi}
\end{eqnarray}
Similarly, the  cyclotron-mode contribution in $S_{\rm eff}^{-}$ is
reproduced from an analogous Lagrangian consisting of a pair of vector
fields $b^{\alpha}_{\mu}=(b^{1}_{\mu},b^{2}_{\mu})$,
\begin{eqnarray}
L_{b}^{\rm cycl}
&=&-A_{\mu}^{(\alpha)} \epsilon^{\mu \nu\rho} \partial_{\nu} b_{\rho}^{\alpha}
- b^{\alpha}_{0} 
+ {1\over{2\rho^{(\alpha)}_{\rm av}}}
\Big[ b^{\alpha} \epsilon \partial b^{\alpha}
+{1\over{\omega_{c}}}\,(b^{\alpha}_{k0})^{2} 
- {1\over{M}}\,(b^{\alpha}_{12})^{2}\Big] ,
\label{Lbminus}
\end{eqnarray}
where summations over layer indices $\alpha$ are understood;
$b^{\alpha} \epsilon \partial b^{\alpha} \equiv b_{\mu}^{\alpha}
\epsilon^{\mu\nu\rho} \partial_{\nu} b_{\rho}^{\alpha}$.
Thus the double-layer system is now described by a gauge-field theory with
the Lagrangian $L^{\rm col}_{\xi}+L_{b}^{\rm cycl}$.

On the other hand, the CS theories, both bosonic and fermionic, lead to an
effective theory of the FQHE, described by the dual-field Lagrangian of
Lee and Zhang~\cite{LZ}.  For the double-layer 
$(m_{1},m_{2},n)$ states it is written in terms of a pair of vector fields
$b^{\alpha}_{\mu}=(b^{1}_{\mu},b^{2}_{\mu})$:
\begin{eqnarray}
L^{\rm CS}_{\rm eff}[b]
&=& - A_{\mu}^{(\alpha)} \epsilon^{\mu \nu\rho} \partial_{\nu}
b_{\rho}^{\alpha} - b^{\alpha}_{0}  +
{1\over{2}}\,\Lambda_{\alpha \beta}\,b^{\alpha} \epsilon
\partial b^{\beta}
+{1\over{2}}\,{1\over{\omega_{c}\rho^{(\alpha)}_{\rm av}}}\,
(b^{\alpha}_{k0})^{2} + \cdots.
\label{LCSeff}
\end{eqnarray}
The mixing matrix  
\begin{eqnarray}
\Lambda_{\alpha \beta}&=&
2\pi \left(\begin{array}{cc}
m_{1} & n\\
  n   & m_{2}
           \end{array} \right)
\label{LCSab}
\end{eqnarray}
is a consequence of the flux attachment characteristic of the
$(m_{1},m_{2},n)$ states.

Our effective Lagrangian $L^{\rm col}_{\xi}+L_{b}^{\rm cycl}$
consists of three vector fields $(\xi_{\mu}, b_{\mu}^{\alpha})$
representing three dipole-active modes.
To make a connection with $L^{\rm CS}_{\rm eff}[b]$ let us try to
rewrite  
$L_{\xi}^{\rm coll}+ L_{b}^{\rm cycl}$ in favor of the new fields
$\eta_{\mu}^{1}=b_{\mu}^{1} +{1\over{2}}\xi_{\mu}$ and
$\eta_{\mu}^{2}=b_{\mu}^{2} -{1\over{2}}\xi_{\mu}$, and eliminate
$\xi_{\mu}$.
We then learn that the resulting Lagrangian $L_{\rm eff}[\eta]$
coincides with  $L^{\rm CS}_{\rm eff}[b]$ to $O(\partial)$ only if 
\begin{equation}
c^{-}={2n\over{m_{1}+m_{2}- 2n}},
\label{cminustwo}
\end{equation}
which thus fixes what appears to be the static structure factor 
of the $(m_{1},m_{2},n)$ state.

While $L_{\rm eff}[\eta]$ and $L^{\rm CS}_{\rm eff}[b]$ coincide to
$O(\partial)$, they differ in $O(\partial^{2})$.
In particular, for the $(m,m,n)$ states they differ merely by the (out-of-phase)
collective-mode spectrum:
\begin{equation}
\epsilon^{-}_{\eta}=  \epsilon_{0}^{-}\,
{m+n\over{2n}} +\cdots \ \ \leftrightarrow \ \ \
\epsilon^{-}_{\rm CS}
={m-n\over{m+n}}\,\omega_{c}.
\label{bainv}
\end{equation}
Here we see spectra of different origin, one of Coulombic origin and one of orbital
origin.

From the above consideration emerge the following observations:
The presence of the dipole-active out-of-phase collective excitations,
inherent to double-layer systems, implies strong interlayer
correlations that affect the long-wavelength characteristics of the
double-layer QH states. The leading $O(\partial)$ correlations are
correctly incorporated into the CS theories by the flux attachment
transformation,  which, however, fails to take in the next-leading
$O(\partial^{2})$ correlations; 
as a result, the collective-excitation spectrum is left on the scale
of $O(\omega_{c})$.

\section{Interlayer coherence and tunneling}

We have so far assumed the absence of interlayer coherence. 
It is possible to extend our analysis to double-layer systems in the
presence of interlayer coherence and tunneling, where a variety of
phenomena such as linearly dispersing  collective modes and Josephson-like
effects~\cite{WZdlayer,EI,SEPW,SGMM,FW} attract attention. The
relevant quantum Hall states are the ground states at filling $\nu
=1/m$ for odd integers $m$, believed to have total pseudospin
$S=N_{e}/2$, with their orbital wave functions well approximated by the
Halperin $(m,m,m)$ wave functions~\cite{BIH}. 

 We have studied the electromagnetic characteristics of the $\nu =1$
double-layer system by means of (i) a low-energy effective theory of
pseudospin textures and  (ii) the single-mode approximation, with
essentially the same results~\cite{KStwo}. 
Here we quote only some of the results.
The low-lying neutral collective mode is a (pseudo) Nambu-Goldstone mode $m(x)$
associated with spontaneous breaking of some pseudospin $U(1)$ symmetry
(exact for vanishing tunneling strength $\triangle_{SAS}\rightarrow 0$).
It is described by an effective Lagrangian of the form
\begin{eqnarray}
{\cal L}^{\rm coll}\!
&=&  {1\over{2}}\, \rho_{s}^{E} \Big[
{1\over{v^{2}}}\,
(\partial_{t} m\! -\!2 {A'}^{-}_{\! 0})^{2} 
-(\partial_{j}m\! -\!2{A'}^{-}_{\!j})^{2} \Big]\!
+{1\over{2}}\,\rho_{\rm av}\triangle_{SAS}\, \cos\, m,
\label{Lcollmode}
\end{eqnarray} 
where $\rho_{s}^{E} \sim O(V^{12}_{\bf p})$ stands for the pseudospin stiffness; 
$v^{2}\approx 2 \rho_{s}^{E}\triangle_{SAS}/\rho_{\rm av}$ and 
$2{A'}^{-}_{\! \mu}  \approx d\,
(\partial_{z}A_{\mu} - \partial_{\mu}A_{z})$ with $d$ being the layer
separation.
This collective mode $m(x)$ gives rise to an electromagnetic response of
the form
\begin{eqnarray}
{\cal L}^{\rm coll}_{\rm em}
&\approx&  {1\over{2}}\,\rho^{E}_{s} d^{2}\,
\Big(\partial_{z}{\bf E}_{\parallel}{\cal D}\partial_{z}{\bf
E}_{\parallel} -
v^{2}\,\partial_{z}B_{\perp}{\cal D}\partial_{z}B_{\perp}\Big)  
\nn&&
+{1\over{4}}\,\rho_{\rm av}\triangle_{SAS}\,d^{2}\,\Big(
E_{\perp}{\cal D} E_{\perp} - v^{2}\, {\bf
B}_{\parallel} {\cal D}{\bf B}_{\parallel} 
\Big)
\end{eqnarray} 
in terms of the field strengths in three space, 
where ${\cal D}=1/\{\omega_{\bf p}^{2}
-(i\partial_{t})^{2}\}$ with the collective-excitation spectrum 
$\omega_{\bf p}^{2} = (\triangle_{SAS})^{2} +\cdots + O(|{\bf p}|)$.
Note that this response is gauge-invariant. 
This implies, in particular, that there is no Anderson-Higgs mechanism or
no Meissner effect working in the present double-layer system.

A close comparison with the effective theory derived within the CS theory reveals that 
the difference is subtle for the $(\partial_{t}m - 2{A'_{0}}^{-})^{2}$ term:
\begin{eqnarray}
&&v^{2}/\rho_{s}^{E}  \leftrightarrow  4\,V^{-}_{{\bf p}=0}
+ 2\triangle_{SAS}/\rho_{\rm av}
\label{coeff}
\end{eqnarray}
These coincide if $V^{-}_{{\bf p}=0}$ reads 
$V^{-}_{{\bf p}=0}
-(1/\rho_{\rm av})\sum_{\bf p}V^{-}_{\bf p}e^{-{1\over{2}}\,{\bf p}^{2}}$;
this shows the importance of Landau-level projection, of which no explicit
account is taken in the CS approach.
For the $(\partial_{j}m - 2A'_{j})^{2}$ term the discrepancy is 
\begin{eqnarray}
\rho_{s}^{E}  \leftrightarrow (\rho_{\rm av}/4M)= \omega_{c}/(8\pi). 
\label{coeff}
\end{eqnarray}
Here we see that the CS approach attributes the pseudospin stiffness
improperly to inter-Landau-level processes.
Another difficulty is that an important low-energy response related to
the drift of Hall electrons, the $A^{-}_{\mu}\epsilon^{\mu\nu\lambda}
\partial_{\nu}A_{\lambda}^{-}$ term, is missing in the CS theory.

All these subtleties derive from the fact that the CS approach, because of the
lack of Landau-level projection, fails to distinguish between the cyclotron
modes and the collective excitation modes. 
The flux attachment in the CS approach properly introduces some crucial
correlations among electrons, but unfortunately not all of them.

\section{Summary and discussion}

In this paper we have studied the electromagnetic characteristics of
double-layer quantum Hall systems.
The results of our analysis are summarized as follows:

\noindent
(1) It is pointed out
that dipole-active excitations, both elementary and
collective, govern the long-wavelength features of quantum Hall (QH)
systems. 
In particular, the presence of the dipole-active
interlayer  out-of-phase collective excitations, inherent to
double-layer systems, modifies the leading $O({\bf k})$ and
$O({\bf k}^{2})$ characteristics 
of the double-layer QH states substantially.

\noindent
(2) The CS theories properly incorporate some crucial correlations
among electrons via the flux-attachment transformation, but not
all of them. 
This explains why they correctly account for
$O(\partial)$ features,
such as the Hall conductance, vortex charges and long-range orders,
while they leave the collective-excitation spectrum on the scale of the
Landau gap$\sim \omega_{c}$.

\noindent
(3) An effective gauge theory reconstructed from such electromagnetic 
characteristics consists of three vector fields representing one
interlayer collective mode and two inter-Landau-level cyclotron modes.
It properly incorporates the single-mode approximation spectrum
of collective excitations, as well as the favorable transport properties
of the standard CS theories.

The idea underlying our approach is to explore the properties of QH
states via their electromagnetic response, which in some cases is
calculable by relying on the incompressible nature of the QH states
without the details of the  microscopic dynamics.  The examples
reported in this paper would combine to enforce again the fact that
incompressibility is the key character of the QH states and prove that
studying the response offers not only a fresh look into the quantum Hall
systems but also a practical means for constructing  effective theories
without referring to composite bosons and fermions. 

\section*{acknowledgments}

This work is supported in part by a Grant-in-Aid for Scientific
Research from the Ministry of Education of Japan,
Science and Culture (No. 14540261).



\section*{References}
\begin{thebibliography}{99}

\bibitem{ZHK} Zhang S C, Hansson T H and Kivelson S 1989 
Phys. Rev. Lett.  62\ \ 82


\bibitem{LF} Lopez A and Fradkin E 1991 Phys. Rev. B 44\ \ 5246

\bibitem{GM} Girvin S M and  MacDonald A H 1987
 Phys. Rev. Lett.  58\ \ 1252

\bibitem{J} Jain J K 1989 Phys. Rev. Lett.  63\ \ 199


\bibitem{MZ}  MacDonald A H and  Zhang S C  1994 Phys. Rev. B 49\ \ 10208

\bibitem{BIH} Halperin B I 1983 Helv. Phys. Acta 56\ \  75

\bibitem{CP} Chakraborty T and Pietilainen P 1987 
 Phys. Rev. Lett.  59\ \ 2784

\bibitem{YMG} Yoshioka D, MacDonald A H and Girvin S M
1989 
 Phys. Rev. B 39 \ \ 1932


\bibitem{GMP} Girvin S M, MacDonald A H and Platzman P M 1986 
Phys. Rev. B 33\ \ 2481

\bibitem{KS} Shizuya K 2002  Phys. Rev. B 65\ \ 205324

\bibitem{FS} Fradkin E and Schaposnik F A 1995 Phys. Lett. B 338\ 235


\bibitem{Moon} 
Moon K, Mori H, Yang K, Girvin S M, MacDonald A H, Zheng L, Yoshioka D and
Zhang S-C 1995 Phys. 

$\!\!\!\!\!\!\!$Rev. B 51\ \ 5138 


\bibitem{RR} Renn S R and Roberts B W 1993 Phys. Rev. B 48\ \ 10926







\bibitem{WZdlayer} 
Wen X G and Zee A 1992 Phys. Rev. Lett. 69\ \ 1811;
1993 Phys. Rev.  B 47\ \  2265

\bibitem{EI} 
Ezawa Z F and Iwazaki A 1992 Int. J. Mod. Phys. B 6\ \  3205;
1993 Phys. Rev.  B 47\ \  7295 

\bibitem{LFdl} Lopez A and Fradkin E  1995 Phys. Rev. B  51\ \  4347

\bibitem{LZ}  Lee D-H and Zhang S-C 1991  Phys. Rev. Lett. 
66\ \ 1220


\bibitem{SEPW} 
Spielman I B,  Eisenstein J P,  Pfeiffer L N and 
West K W 2001  Phys. Rev. Lett. 87\ \ 036803 


\bibitem{SGMM}
Stern A, Girvin S M, MacDonald A H, Ma N 2001 
Phys. Rev. Lett. 86\ \  1829 

\bibitem{FW}
Fogler M M and Wilczek F 2001  Phys. Rev. Lett. 86\ \ 1833


\bibitem{KStwo} Shizuya K in preparation


\end{thebibliography}
\end{document}